\documentclass[pra,twocolumn,showpacs]{revtex4}
\usepackage[usenames]{color}
\usepackage{epsf}
\usepackage{epsfig}
\usepackage{graphicx}
\usepackage{latexsym}
\usepackage{bm}
\usepackage{color}
\usepackage{amsmath}
\usepackage[normalem]{ulem}
\usepackage{todonotes}

\begin{document}
Phys. Rev. A 90, 022515 (2014)

\title{Interatomic-Coulombic-Decay induced recapture of photoelectrons in Helium Dimers}

\author{P. Burzynski$^{1}$} 
\author{F. Trinter$^{1}$}
\author{J. B. Williams$^{1}$}
\author{M. Weller$^{1}$}
\author{M. Waitz$^{1}$}
\author{M. Pitzer$^{1}$}
\author{J. Voigtsberger$^{1}$}
\author{C. Schober$^{1}$}
\author{G. Kastirke$^{1}$}
\author{C. M\"uller$^{1}$}
\author{C. Goihl$^{1}$}
\author{F. Wiegandt$^{1}$}
\author{R. Wallauer$^{1}$}
\author{A. Kalinin$^{1}$}
\author{L. Ph. H. Schmidt$^{1}$}
\author{M. Sch\"offler$^{1}$}
\author{G. Schiwietz$^{2}$}
\author{N. Sisourat$^{3,4}$}
\author{T. Jahnke$^{1}$}
\author{R. D\"orner$^{1}$}

\email {doerner@atom.uni-frankfurt.de}

\address{
$^1$ Institut f\"ur Kernphysik, J.~W.~Goethe-Universit\"at, Max-von-Laue-Str. 1, 60438 Frankfurt am Main, Germany \\
}
\address{
$^2$ Helmholtz-Zentrum Berlin f. Materialien u. Energie, Institute G-ISRR, Hahn-Meitner-Platz 1, 14109 Berlin, Germany \\
}
\address{
$^3$ Sorbonne Universit\'es, UMR 7614, Laboratoire de Chimie Physique Mati\`ere et Rayonnement, F-75005 Paris, France \\
}
\address{
$^4$ CNRS, UMR 7614, Laboratoire de Chimie Physique Mati\`ere et Rayonnement, F-75005 Paris, France \\
}

\begin{abstract}

We investigate the onset of photoionization shakeup induced interatomic Coulombic decay (ICD) in He$_2$ at the He$^{+*}(n=2)$ threshold by detecting two He$^+$ ions in coincidence. We find this threshold to be shifted towards higher energies compared to the same threshold in the monomer. The shifted onset of ion pairs created by ICD is attributed to a recapture of the threshold photoelectron after the emission of the faster ICD electron. 
\end{abstract}

\maketitle

Excited ions can get rid of their excess energy via the emission of a photon or an electron. If, however, the excited atom is spatially close to other atoms and the excitation energy is above the ionization threshold of this neighbor,  the excess energy can also be transfered to the neighbor where it leads to emission of an electron. This energy transfer process is termed interatomic Coulombic decay (ICD). It was introduced by Cederbaum and coworkers in 1997 \cite{Cederbaum97prl} and was demonstrated experimentally first for Neon clusters \cite{marburger2003prl} and Neon dimers \cite{Jahnke04icdprl}. The related interatomic Auger transitions in solid matter have ofen been discussed, but broad valence bands, surface/bulk differences and significant electron energy-loss processes do typically preclude a clear assignment of this process for solids. Many studies have shown since then that ICD is a very general phenomenon occuring in van der Waals bound (see e.g. \cite{marburger2003prl, Jahnke04icdprl, ohrwall2004prl, Ouchi11_2}) and hydrogen bound systems (see e.g. \cite{Mueller06,Jahnke10,Mucke10}). It can be induced by photoionization (see e.g. \cite{marburger2003prl,Jahnke04icdprl}), photoexcitation \cite{Trinter13b,golan2012jpcl,Gokhberg2005epl,Najjari2010prl}, Auger decay \cite{Santra03,Kreidi08_2}, ion impact \cite{Kim11,Kim13,Titze11}, and electron impact \cite{Yan13} or as in the present case after shakeup \cite{Jahnke07}. The most extreme system for which ICD has been reported is the Helium dimer \cite{Havermeier10_1,Sisourat10_1}. The neutral He dimer is very weakly bound (about 95 neV) and the internuclear distance extends to very large distances, with the mean distance of about 52 \AA. ICD in He$_2$ can occur when one of the Helium atoms is ionized and its remaining electron is shaken up to any excited state (He$^{+*}$(n=2,3...)). In the next step, the He$^{+*}$(n=2)He contracts and during that nuclear motion it undergoes ICD. The electron of the exited He$^+$ relaxes to the ground state and the energy is transfered to the neutral neighbor where the ICD electron is emitted. Finally, the two He$^+$ ions Coulomb explode back-to-back:

He-He $\xrightarrow{h\nu}$ He$^{+*}$-He + e$_{ph} \xrightarrow{ICD}$ He$^+ +$ He$^+$ +  e$_{ph}$ + e$_{ICD}$

Usually, ICD and the subsequent Coulomb explosion is discussed in a two step picture, where the decay is independent from the initial ionization/excitation process. In the present work, we show that close to the ionization/excitation threshold this two step approximation breaks down.  Photoelectron and ICD electron interaction can lead to recapture of the photoelectron into a bound state of one of the two ions, which quenches the Coulomb explosion. A direct link between the ionization process and ICD has been discussed in two contexts in the literature so far. The first is the recoil effect, where it has been shown experimentally \cite{Kreidi09} and theoretically \cite{Demekhin09_2} that the recoil momentum of the photoelectron or an Auger electron can induce nuclear motion, which in turn modifies the ICD energy spectrum. The second context is closely related to the present work, where Trinter et al. \cite{Trinter13a} have seen a shift of the photoelectron energy due to post collision interaction (PCI) with the ICD electron \cite{Russek86,Landers09prl}. In a time dependent picture the photoelectron is originally created in the potential of a singly charged species. After some delay the ICD electron is emitted, but then it surpasses the slow photoelectron, which from then on feels the attractive potential of a doubly charged ion. This slows down the photoelectron. This streaking towards lower photoelectron energies depends on the time delay between the photoabsorption and the ICD electron and can easily be modeled. It has recently been used to make the first  'movie' of nuclear motion during ICD \cite{Trinter13a}. The same process of post collision interaction leads, for very small photoelectron energies, to a recapture of a part of the photoelectron wave packet, which is the effect we study here. For atomic Auger decay following innershell ionization, PCI is well understood \cite{Russek86,Landers09prl}. Recently Sch\"utte et al. \cite{schuette12prl} have experimentally verified the time dependent picture we have just discussed, by showing that the energy exchange between photoelectron and Auger electron indeed depends on the Auger emission time delay. For atomic multiple ionization, such recapture of the photoelectron by PCI, leads to a shifted onset of the production of higher charge states, as e.g. reported for Ar \cite{samsonpra1996,Guillemin12prl}.

The present experiment has been performed at beam line UE112-PGM-1 in the synchrotron radiation facility BESSY (Berlin) during single bunch operation using a COLTRIMS reaction microscope \cite{Ullrich03rpp,doerner00pr,Jahnke04}. The photon beam was intersected with a supersonic He gas jet in the center of the COLTRIMS spectrometer. A 7.5 V/cm homogeneous electric field guided the ions towards a position sensitive micro channel plate detector with hexagonal delay-line readout (RoentDek HEX90) \cite{Jagutzki02ieee}. A nozzle temperature of 21 K at 2.5 bar driving pressure resulted in a fraction of about 1-2\% He$_2$ in the atomic gas jet. The photon energy was scanned across the He$^+$(n=2) threshold. In the offline analysis the momentum vectors of the ions were obtained from the position of impact at the detector and the time-of-flight. We analyzed ion pairs emitted back-to-back and the simultaneously measured He$^+$(n=2) monomer ions from the atomic helium fraction of the gas jet. The back-to-back events dominate the recoil pair-correlation function slightly above threshold. This fact is related to a small fraction of sequential pair production due to the limited photon density of the energy-filtered synchrotron beam, to an extremly small probability of the correlated electron knock-off transitions close to the threshold and to a small fraction of secondary ion/atom collisions in the gas jet. He$^+$(n=2) ions have been discriminated from the He$^+$(n=1) ions by the ion momentum vector (see \cite{Doerner96ratio} Fig. 1). Close to the He$^+(n=2)$ threshold the He$^+$(n=1) ions carry the 1.7 a.u. recoil momentum of the photoelectron while the He$^{+*}$(n=2) ions are accompanied by a zero kinetic energy photoelectron and hence have almost no recoil momentum.

\begin{figure}[h!t]
  \begin{center}
  \includegraphics[width=7. cm]{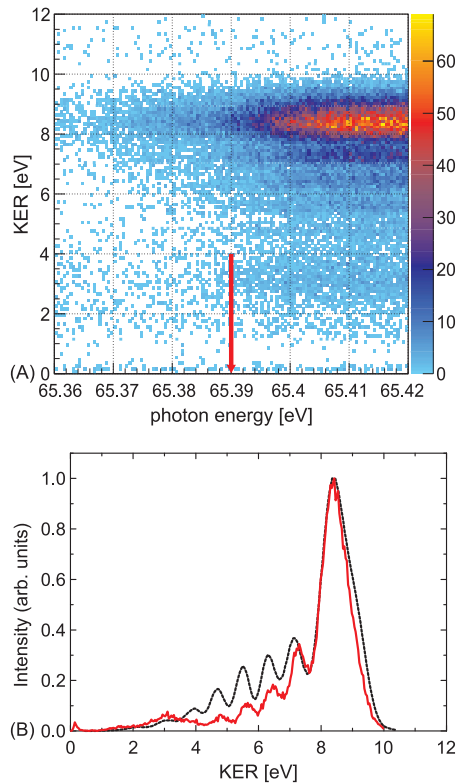}
  \caption{(A) Kinetic energy release (KER) of He$^+$ ion pairs as function of photon energy, red arrow shows the threshold for creating He$^{+*}$(n=2) for a helium atom. (B) Red full line: Projection of (A) onto the y-axis for photon energy range 65.41-65.42 eV. Black dotted line: Theoretical KER distribution. The two curves are normalized to the maximum.}
  \end{center}
\end{figure}

Fig. 1  shows the measured kinetic energy release (KER, summed over both recoil-ion energies) as function of the photon energy. The KER above the He$^{+*}$(n=2) threshold (E$_\gamma$ range=65.41-65.42 eV, Fig. 1b) is in excellent agreement with published work \cite{Havermeier10_1}. It shows a vibrational structure from the contracting dimer with a maximum between 8-9.5 eV, which results from nuclear wave packet hitting the inner turning point on the potential energy surface of the excited dimer ion (see \cite{Sisourat10_1,Sisourat10_2} for a detailed analysis). 


\begin{figure}[ht]
  \begin{center}
  \includegraphics[width=7.6 cm]{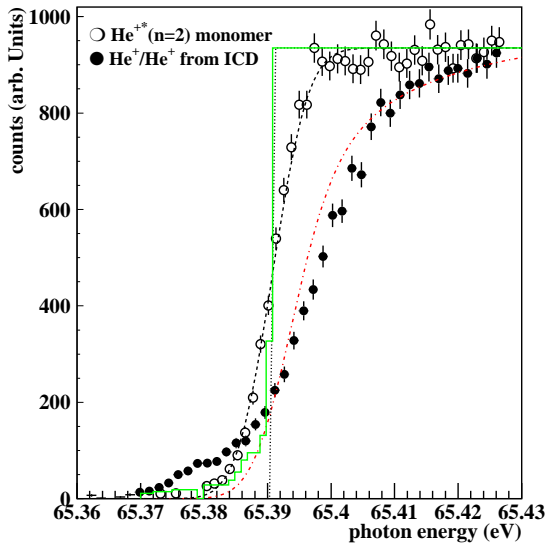}
  \caption{Photon energy dependence of the count rate for:  two He$^+$ ions (filled circles) (projection of data in Fig. 1a onto the x-axis) and He$^{+*}$(n=2) ions from ionization of the monomer (open circles). Black dotted step function: threshold for He$^{+*}$(n=2), black dashed line: step function convoluted with a Gaussian function ($\sigma$ = 4.5 meV) for the energy resolution of the beam line. Green full line: expected onset of ICD without taking PCI into account (see text), red dash dotted line: model including recapture of photoelectrons by PCI using calculated widths from Table I (see text) as well as the experimental resolution of $\sigma$= 4.5 meV. The data points are normalized at the highest points.
  }
  \end{center}
\end{figure}


Fig. 2 shows the photon energy dependence of the ion pair count rate, i.e. a projection of the data from Fig. 1a onto the horizontal axis (filled circles) and count rate from $He^{+*}$(n=2) monomers (open circles). Both datasets are normalized to the highest energy point. A constant background for the ion pairs probably resulting from a knock off  process \cite{Havermeier10prl3,Ni2013pra} and higher harmonics from  the beam line  has been subtracted. 

Owing to the resolution of the beam line, the count rates are not a step function at the He$^{+*}$(n=2) threshold. The resolution of the beam line is $\sigma$ = 4.5 meV which is extracted from the data of the monomers. These monomers come from atomic helium. The black dashed line of Fig. 2 is a convolution of the step function (black dotted) with a Gaussian function with a FWHM of 4.5 meV. This line follows very well the monomer ion count rate. The ion pair count rate shows a significant energy offset compared to the monomers. Note that monomers and ion pairs are measured simultaneous in our setup, which excludes any possible systematic error on the photon energy as the origin of this energy offset. We will show now that the shifted onset of the ion pair production is caused by recapture of the slow near threshold photoelectron after ICD, which neutralizes one of the two ions. A classical modeling of this recapture process, summing over all calculated ICD channels using calculated ICD lifetimes, results in the dash dotted red line in Fig. 2, which nicely reproduces the shifted onset. There are no free parameters in this calculation. The model is presented in the following.

In the present data, the photoelectron energy is below 50 meV, while the ICD electron has an energy of 6-15 eV. 
Even though there is a time delay ($t_{ICD}$) between the emission of the photoelectron and ICD electron, the ICD electron will always overtake the photoelectron in the vicinity of the residual ion(s), because $t_{ICD}$ is small and the ICD electron is much faster than the photoelectron. 

 Note that $t_{ICD}$ is not the lifetime of the respective ICD channel, but the time at which ICD occurs for this individual event. For the present case the travel time of the ICD electron is negligible compared to $t_{ICD}$. The photoelectron therefore starts its way to the continuum initially leaving behind a singly charged species. 
 However, when the ICD electron is emitted the ion charge state increases from singly to doubly charged, which changes the potential the photoeletron feels, at least after it has been overtaken by the ICD electron, from $-1/r_e$ to $-2/r_e$ (in atomic units), where  $r_e$ is the distance the photoelectron has reached at time $t_{ICD}$. 

We calculate the trajectory of the photoelectron classically by starting the electron at a distance $r_{es}$ in a Coulomb potential with an initial kinetic energy $E_\gamma-E_{IPX}+1/r_{es}$ where $E_{\gamma}$ is the photon energy and $E_{IPX}$ is the ionization potential of He plus the energy it takes to excite He$^+$ from its ground state to $He^{+*}$(n=2). We have chosen $r_{es}=10$ a.u. and have verified that the results are insensitive of this choice over a wide range. Without the ICD electron the photoelectron would escape to the continuum. However, if ICD occurs, the electron loses the energy  $-1/r_e$ and might get trapped in the ionic potential.  
For every particular electronic and vibrational He$^{+*}$(n=2)He state ICD occurs with an exponential time dependence. We calculate the fraction of recapture photoelectrons using this exponential distribution of $t_{ICD}$ for each photon energy. We then sum over all electronic and vibrational states. The electronic states are weighted with their statistical weight and the vibrational states with their Franck-Condon overlap with the $He_2$ ground state. Franck-Condon factors, energies and ICD lifetimes for each state are given in Table I. 
\begin{table}[ht]
\begin{center}
\begin{tabular} {|c|c|c|c|c|} \hline
$\nu$  & $^2\Sigma_g^+ : 2p_z, 2s$ &$^2\Sigma_u^+ : 2p_z, 2s$  &$^2\Pi_g:2p_{x,y}$&$^2\Pi_u:2p_{x,y}$ \\ \hline
\multicolumn{5}{|c|}{\bfseries Franck-Condon factors} \\ \hline
0  & $7.25\cdot 10^{-5}$ & $2.95\cdot 10^{-5}$ & $5.34\cdot 10^{-6}$& $1.70\cdot 10^{-5}$ \\ \hline
1  & $3.85\cdot 10^{-4}$ & $2.11\cdot 10^{-4}$ & $5.47\cdot 10^{-5}$& $1.39\cdot 10^{-4}$ \\ \hline
2  & $1.55\cdot 10^{-3}$ & $9.40\cdot 10^{-4}$ & $3.15\cdot 10^{-4}$& $6.80\cdot 10^{-4}$ \\ \hline
3  & $4.56\cdot 10^{-3}$ & $2.91\cdot 10^{-3}$ & $1.29\cdot 10^{-3}$& $2.39\cdot 10^{-3}$ \\ \hline
4  & $7.31\cdot 10^{-3}$ & $6.20\cdot 10^{-3}$ & $4.34\cdot 10^{-3}$& $7.26\cdot 10^{-3}$ \\ \hline
5  & $3.30\cdot 10^{-2}$ & $2.10\cdot 10^{-2}$ & $1.26\cdot 10^{-2}$& $2.06\cdot 10^{-2}$ \\ \hline
6  & $2.72\cdot 10^{-2}$ & $1.82\cdot 10^{-2}$ & $3.16\cdot 10^{-2}$& $5.30\cdot 10^{-2}$ \\ \hline
7  & $4.92\cdot 10^{-2}$ & $1.36\cdot 10^{-1}$ & $8.86\cdot 10^{-2}$& $1.88\cdot 10^{-1}$ \\ \hline
8  & $3.13\cdot 10^{-1}$ & $1.20\cdot 10^{-1}$ & $3.88\cdot 10^{-1}$& - \\ \hline
\multicolumn{5}{|c|}{\bfseries lifetimes [fs]} \\ \hline
0  & 30.61 & 33.49 & 34.08 & 22.00 \\ \hline
1  & 37.40 & 36.44 & 38.25 & 27.15 \\ \hline
2  & 51.89 & 49.00 & 47.43 & 39.27 \\ \hline
3  & 81.07 & 69.58 & 66.06 & 61.22 \\ \hline
4  & 147.10 & 113.76 & 111.16 & 127.44 \\ \hline
5  & 377.19 & 261.05 & 251.72 & 387.04 \\ \hline
6  & 1420 & 957.02 & 839.12 & 1810 \\ \hline
7  & 1975 & 2640 & 4560 & 15570 \\ \hline
8  & 24690 & 8710 & 53960 & - \\ \hline
\multicolumn{5}{|c|}{\begin{tabular}{c} \bfseries negative shifts of the vibrational level\\ \bfseries compared to the monomer [eV]\end{tabular}} \\ \hline
0  & 1.24$\cdot 10^{-1}$ & 1.45$\cdot 10^{-1}$ & 1.68$\cdot 10^{-1}$ & 1.43$\cdot 10^{-1}$ \\ \hline
1  & 8.01$\cdot 10^{-2}$ & 9.36$\cdot 10^{-2}$ & 1.13$\cdot 10^{-1}$ & 9.09$\cdot 10^{-2}$ \\ \hline
2  & 4.61$\cdot 10^{-2}$ & 5.72$\cdot 10^{-2}$ & 6.95$\cdot 10^{-2}$ & 5.19$\cdot 10^{-2}$ \\ \hline
3  & 2.38$\cdot 10^{-2}$ & 3.09$\cdot 10^{-2}$ & 3.79$\cdot 10^{-2}$ & 2.58$\cdot 10^{-2}$ \\ \hline
4  & 1.06$\cdot 10^{-2}$ & 1.45$\cdot 10^{-2}$ & 1.74$\cdot 10^{-2}$ & 1.07$\cdot 10^{-2}$ \\ \hline
5  & 4.13$\cdot 10^{-3}$ & 5.94$\cdot 10^{-3}$ & 6.40$\cdot 10^{-3}$ & 3.34$\cdot 10^{-3}$ \\ \hline
6  & 9.09$\cdot 10^{-4}$ & 1.50$\cdot 10^{-3}$ & 1.70$\cdot 10^{-3}$ & 6.75$\cdot 10^{-4}$ \\ \hline
7  & 7.03$\cdot 10^{-5}$ & 3.75$\cdot 10^{-4}$ & 2.35$\cdot 10^{-4}$ & 4.37$\cdot 10^{-5}$ \\ \hline
8  & 5.35$\cdot 10^{-5}$ & 1.48$\cdot 10^{-5}$ & 2.99$\cdot 10^{-6}$ & - \\ \hline
\multicolumn{5}{|c|}{ dissociation limit 65.393 eV} \\ \hline
\end{tabular}
\caption{Calculated characteristics of the He$^{+*}$(n=2)He states: Franck-Condon factors for the overlap with the He$_2$ ground state, vibrational state energies and ICD lifetimes \cite{Sisourat10_2}.}
\end{center}
\end{table}

There are two counteracting effects included in this calculation. First the threshold for each state  He$^{+*}$(n=2)He will be slightly below the He$^{+*}$(n=2) threshold for the monomer, because of the larger size of the combined electronic potential due to both heavy constituents of the dimer (see Table I for binding energies). Thus neglecting the recapture the threshold for ion pair creation would be slightly smeared to lower energies,  as shown by the full green curve in Fig. 2. If one includes the recapture as described above the full green curve is modified to the dash dotted red curve in Fig. 2, which is shifted towards higher photon energies. This dash dotted red curve also includes the photon energy resolution which we obtained in situ from the monomer. This calculation captures the main effect seen in the experiment. Experimentally, we also see a small contribution of ion pairs below threshold,  which is not reproduced by our calculation. The origin of these contributions below threshold is unclear, but might be related to details of the angular distribution of ejected ICD and photoelectrons or to the accuracy of the computed negative energy shift.
     
In conclusion, we have shown that close to threshold fragment creation by ICD cannot be treated independently of the ionization process. Post collision interaction of the photoelectron with the ICD electron can even lead to recapture of the photoelectron. This observed effect is similar to PCI between photoelectron and fast Auger electrons in atomic species. While we have studied only the monopole term of this post collision here, we expect that the angular distributions of the electrons will also be altered, an effect studied recently for the atomic Auger case \cite{Landers09prl}.  The discussed recapture will also be active at the threshold for creation of He$^{+*}$ in n=3 and higher. In this case the angular distributions of photoelectron and ICD electron will be different(see \cite{havermeier2010}). We believe that PCI in the continuum will in the future become a major tool for ultrafast time resolved studies, as shown recently in pioneering work by Trinter et al. \cite{Trinter13a}.

\acknowledgments   We thank the staff of BESSY II for
experimental support. This work was funded by the Deutsche Forschungsgemeinschaft and supported by RoentDek Handels GmbH.

\bibliographystyle{unsrt}

\begin{thebibliography}{10}

\bibitem{Cederbaum97prl}
L.S. Cederbaum, J. Zobeley and F. Tarantelli,
\newblock {\em Phys. Rev. Lett.} \textbf{79,} 4778 (1997).

\bibitem{marburger2003prl}
S. Marburger, O. Kugeler, U. Hergenhahn and T. M\"oller,
\newblock {\em Phys. Rev. Lett.} \textbf{90,} 203401 (2003).

\bibitem{Jahnke04icdprl}
T. Jahnke, A. Czasch, M. S. Sch\"offler, S. Sch\"ossler, A. Knapp, M. K\"asz, J. Titze, C. Wimmer, K. Kreidi, R. E. Grisenti, A. Staudte, O. Jagutzki, U. Hergenhahn, H. Schmidt-B\"ocking  and R. D\"{o}rner,
\newblock {\em Phys. Rev. Lett.} \textbf{93,} 163401 (2004).

\bibitem{ohrwall2004prl}
G. \"Ohrwall, M. Tchaplyguine, M. Lundwall, R. Feifel, H. Bergersen, T. Rander, A. Lindblad, J. Schulz, S. Peredkov, S. Barth, S. Marburger, U. Hergenhahn, S. Svensson and O. Bj\"orneholm,
\newblock {\em Phys. Rev. Lett.} \textbf{93,} 173401 (2004).

\bibitem{Ouchi11_2}
T. Ouchi, K. Sakai, H. Fukuzawa, X.-J. Liu, I. Higuchi, Y. Tamenori, K. Nagaya, H. Iwayama, M. Yao, D. Zhang, D. Ding, A. I. Kuleff, S. D. Stoychev, Ph. V. Demekhin, N. Saito and K. Ueda,
\newblock {\em Phys. Rev. Lett.} \textbf{107,} 053401 (2011).

\bibitem{Mueller06}
I. B. M\"uller and L. S. Cederbaum,
\newblock {\em J. Chem. Phys.} \textbf{125,} 204305 (2006).

\bibitem{Jahnke10}
T. Jahnke, H. Sann, T. Havermeier, K. Kreidi, C. Stuck, M. Meckel, M. Sch\"offler, N. Neumann, R.Wallauer, S. Voss, A.  Czasch, O. Jagutzki, A. Malakzadeh, F. Afaneh, Th. Weber, H. Schmidt-B\"ocking and R. D\"orner,
\newblock {\em Nature Physics} \textbf{6,} 139 (2010).

\bibitem{Mucke10}
M. Mucke, M. Braune, S. Barth, M. F\"orstel, T. Lischke, V. Ulrich, T. Arion, U. Becker, A. Bradshaw and U. Hergenhahn,
\newblock {\em Nature Physics} \textbf{6,} 143 (2010).

\bibitem{Trinter13b}
F. Trinter, J. B. Williams, M. Weller, M. Waitz, M. Pitzer, J. Voigtsberger, C. Schober, G. Kastirke, C. M\"uller, C. Goihl, P. Burzynski, F. Wiegandt, R. Wallauer, A. Kalinin, L. Ph. H. Schmidt, M. Sch\"offler, Y.-C. Chiang, K. Gokhberg, T. Jahnke, R. D\"orner,
\newblock {\em Phys. Rev. Lett.} \textbf{111,} 233004 (2013).

\bibitem{golan2012jpcl}
A. Golan and M. Ahmed,
\newblock {\em J. Phys. Chem. Lett.} \textbf{3,} 458 (2012).

\bibitem{Gokhberg2005epl}
K. Gokhberg, A. B. Trofimov, T. Sommerfeld and L. S. Cederbaum,
\newblock {\em Europhys. Lett.} \textbf{72(2),} 228 (2005).

\bibitem{Najjari2010prl}
B. Najjari, A.B. Voitkiv. and C. M\"uller,
\newblock {\em Phys. Rev. Lett.} \textbf{105,} 153002 (2010).

\bibitem{Santra03}
R. Santra and L. S. Cederbaum,
\newblock {\em Phys. Rev. Lett.} \textbf{90,} 153401 (2003).

\bibitem{Kreidi08_2}
K. Kreidi, T. Jahnke, Th. Weber, T. Havermeier, X. Liu, Y. Morisita, S. Sch\"ossler, L. Ph. H. Schmidt, M. Sch\"offler, M. Odenweller, N. Neumann, L. Foucar, J. Titze, B. Ulrich, F. Sturm, C. Stuck, R. Wallauer, S. Voss, I. Lauter, H. K. Kim, M. Rudloff, H. Fukuzawa, G. Pr\"umper, N. Saito, K. Ueda, A. Czasch, O. Jagutzki, H. Schmidt-B\"ocking, S. Stoychev, Ph. V. Demekhin and R. D\"orner,
\newblock {\em Phys. Rev. A.} \textbf{78,} 043422 (2008).

\bibitem{Kim11}
H. K. Kim, J. Titze, M. Sch\"offler, F. Trinter, M. Waitz, J. Voigtsberger, H. Sann, M. Meckel, Ch. Stuck, U. Lenz, M. Odenweller, N. Neumann, S. Sch\"ossler, K. Ullmann-Pfleger, B. Ulrich, R. C. Fraga, N. Petridis, D. Metz, A. Jung, R. Grisenti, A. Czasch, O. Jagutzki, L. Schmidt, T. Jahnke, H. Schmidt-B\"ocking and R. D\"orner,
\newblock {\em PNAS} \textbf{108,} 11821 (2011).

\bibitem{Kim13}
H. K. Kim, H. Gassert, M. Sch\"offler, J. Titze, M. Waitz, J. Voigtsberger, F. Trinter, J. Becht, A. Kalinin, N. Neumann, C. Zhou, L. Schmidt, O. Jagutzki, A. Czasch, H. Merabet, H. Schmidt-B\"ocking, T. Jahnke, A. Cassimi and R. D\"orner,
\newblock {\em Phys. Rev. A.} \textbf{88,} 042707 (2013).

\bibitem{Titze11}
J. Titze, M. Sch\"offler, H. -K. Kim, F. Trinter, M. Waitz, J. Voigtsberger, N. Neumann, B. Ulrich, K. Kreidi, R. Wallauer, M. Odenweller, T. Havermeier, S. Sch\"ossler, M. Meckel, L. Foucar, T. Jahnke, A. Czasch, L. Schmidt, O. Jagutzki, R. E. Grisenti, H. Schmidt-B\"ocking, H. J. L\"udde and R. D\"orner,
\newblock {\em Phys. Rev. Lett.} \textbf{106,} 033201 (2011).

\bibitem{Yan13}
S. Yan, P. Zhang, X. Ma, S. Xu, B. Li, X. L. Zhu, W. T. Feng, S. F. Zhang, D. M. Zhao, R. Zhang, D. Guo and H. P. Liu,
\newblock {\em Phys. Rev. A.} \textbf{88,} 042712 (2013).

\bibitem{Jahnke07}
T. Jahnke, A. Czasch, M. Sch\"offler, S. Sch\"ossler, M. K\"asz, J. Titze, K. Kreidi, R. E. Grisenti, A. Staudte, O. Jagutzki, L. Schmidt, Th. Weber, H. Schmidt-B\"ocking, K. Ueda and R. D\"orner,
\newblock {\em Phys. Rev. Lett.} \textbf{99,} 153401 (2007).

\bibitem{Havermeier10_1} 
T. Havermeier, T. Jahnke, K. Kreidi, R. Wallauer, S. Voss, M. Sch\"offler, S. Sch\"ossler, L. Foucar, N. Neumann, J. Titze, H. Sann, M. K\"uhnel, J. Voigtsberger, J. Morilla, W. Sch\"ollkopf, H. Schmidt-B\"ocking, R. Grisenti and R. D\"orner,
\newblock {\em Phys. Rev. Lett.} \textbf{104,} 133401 (2010).

\bibitem{Sisourat10_1}
N. Sisourat, N. Kryzhevoi, P. Kolorenc, S. Scheit, T. Jahnke and L. Cederbaum,
\newblock {\em Nature Physics} \textbf{6,} 508 (2010).

\bibitem{Kreidi09}
K. Kreidi, Ph. Demekhin, T. Jahnke, Th. Weber, T. Havermeier, X. Liu, Y. Morisita, S. Sch\"ossler, L. Schmidt, M. Sch\"offler, M. Odenweller, N. Neumann, L. Foucar, J. Titze, B. Ulrich, F. Sturm, C. Stuck, R. Wallauer, S. Voss, I. Lauter, H. Kim, M. Rudloff, H. Fukuzawa, G. Pr\"umper, N. Saito, K. Ueda, A. Czasch, O. Jagutzki, H. Schmidt-B\"ocking, S. Scheit, L. Cederbaum and R. D\"orner,
\newblock {\em Phys. Rev. Lett.} \textbf{103,} 033001 (2009).

\bibitem{Demekhin09_2}
Ph. Demekhin, S. Scheit and L. Cederbaum,
\newblock {\em J. Chem. Phys.} \textbf{131,} 164301 (2009).

\bibitem{Trinter13a}
F. Trinter, J. Williams, M. Weller, M. Waitz, M. Pitzer, J. Voigtsberger, C. Schober, G. Kastirke, C. M\"uller, C. Goihl, P. Burzynski, F. Wiegandt, T. Bauer, R. Wallauer, H. Sann, A. Kalinin, L. Schmidt, M. Sch\"offler, N. Sisourat and T. Jahnke,
\newblock {\em Phys. Rev. Lett.} \textbf{111,} 093401 (2013).

\bibitem{Russek86}
A. Russek and W. Mehlhorn,
\newblock {\em J. Phys. B: At., Mol. Opt. Phys.} \textbf{19,} 911-927 (1986).

\bibitem{Landers09prl}
A. Landers, F. Robicheaux, T. Jahnke, M. Sch\"offler, T. Osipov, J. Titze, S. Lee, H. Adaniya, M. Hertlein, P. Ranitovic, I. Bocharova, D. Akoury, A. Bhandary, Th. Weber, M. Prior, C. Cocke, R. D\"orner and A. Belkacem,
\newblock {\em Phys. Rev. Lett.} \textbf{102,} 223001 (2009).

\bibitem{schuette12prl}
B. Sch\"utte, S. Bauch, U. Fr\"uhling, M. Wieland, M. Gensch, E. Pl\"onjes, T. Gaumnitz, A. Azima, M. Bonitz and M. Drescher,
\newblock {\em Phys. Rev. Lett.} \textbf{108,} 253003 (2012).

\bibitem{samsonpra1996}
J. Samson, W. Stolte, Z. He, J. Cutler and D. Hansen,
\newblock {\em Phys. Rev. A.} \textbf{54,} 2099-2106 (1996).

\bibitem{Guillemin12prl}
R. Guillemin, S.  Sheinerman, C. Bomme, L. Journel, T. Marin, T. Marchenko, R. K. Kushawaha, N. Trcera, M. N. Piancastelli and M. Simon,
\newblock {\em Phys. Rev. Lett.} \textbf{109,} 013001 (2012).

\bibitem{Ullrich03rpp}
J. Ullrich, R. Moshammer, A. Dorn, R. D\"orner, L. Ph. Schmidt and H. Schmidt-B\"ocking,
\newblock {\em Rep. Prog. Phys.} \textbf{66,} 1463-1545 (2003).

\bibitem{doerner00pr}
R. D\"orner, V. Mergel, O. Jagutzki, L. Spielberger, J. Ullrich, R. Moshammer and H. Schmidt-B\"ocking,
\newblock {\em Physics Reports} \textbf{330,} 96-192 (2000).

\bibitem{Jahnke04}
T. Jahnke, Th. Weber, T. Osipov, A. L. Landers, O. Jagutzki, L. Ph. H. Schmidt, C. L. Cocke, M. H. Prior, H. Schmidt-B\"ocking and R. D\"orner,
\newblock {\em J. Electron Spectrosc. Relelat. Phenom.} \textbf{141,} 229 (2004).

\bibitem{Jagutzki02ieee}
O. Jagutzki, A. Cerezo, A. Czasch, R. D\"orner, M. Hattass, M. Huang, V. Mergel, U. Spillmann, K. Ullmann-Pfleger, Th. Weber, H. Schmidt-B\"ocking and G.D.W. Smith,
\newblock {\em IEEE Transact. on Nucl. Science} \textbf{49,} 2477 (2002).

\bibitem{Doerner96ratio}
R. D\"orner, T. Vogt, V. Mergel, H. Khemliche, S. Kravis, C.L. Cocke, J. Ullrich, M. Unverzagt, L. Spielberger, M. Damrau, O. Jagutzki, I. Ali, B. Weaver, K. Ullmann, C.C. Hsu, M. Jung, E.P. Kanter, B. Sonntag, M.H. Prior, E. Rotenberg, J. Denlinger,    T. Warwick, S.T. Manson and H. Schmidt-B\"ocking,
\newblock {\em Phys. Rev. Lett.} \textbf{76,} 2654 (1996).

\bibitem{Sisourat10_2}
N. Sisourat, N. V. Kryzhevoi, P. Kolorenc, S. Scheit and L. S. Cederbaum,
\newblock {\em Phys. Rev. A.} \textbf{82,} 053401 (2010).

\bibitem{Havermeier10prl3}
T. Havermeier, T. Jahnke, K. Kreidi, R. Wallauer, S. Voss, M. Sch\"offler, S. Sch\"ossler, L. Foucar, N. Neumann, J. Titze, H. Sann, M. Kuehnel, J. Voigtsberger, A. Malakzadeh, N. Sisourat, W. Sch\"ollkopf, H. Schmidt-B\"ocking, R. E. Grisenti and R. D\"orner,
\newblock {\em Phys. Rev. Lett.} \textbf{104(15),} 153401 (2010).

\bibitem{Ni2013pra}
H. Ni, C. Ruiz, R. D\"orner and A. Becker,
\newblock {\em Phys. Rev. A.} \textbf{88,} 013407 (2013).

\bibitem{havermeier2010}
T. Havermeier, K. Kreidi, R. Wallauer, S. Voss, M. Sch\"offler, S. Sch\"ossler, L. Foucar, N. Neumann, J. Titze, H. Sann, M. K\"uhnel, J. Voigsbeger, N. Sisourat, W. Sch\"ollkopf, H. Schmidt-B\"ocking, R. E. Gristenti, R. D\"orner and T. Jahnke,
\newblock {\em Phys. Rev. A.} \textbf{82,} 063405 (2010).



\end{thebibliography}

\end{document}